\documentclass[12pt,preprint]{aastex}
\usepackage[]{natbib}

\shorttitle{Variability of CSCha}
\shortauthors{Nagel et al.}

\begin{document}
\bibliographystyle{plainnat}

\title{MIR variability of the binary system CSCha}

\author{Erick Nagel\altaffilmark{1},  Catherine Espaillat\altaffilmark{2,5},
Paola D'Alessio\altaffilmark{3}, and Nuria Calvet\altaffilmark{4}}
\affil{\altaffilmark{1}Departamento de Astronom\'\i a, Universidad de 
Guanajuato, Guanajuato, Gto, M\'exico 36240}
\affil{\altaffilmark{2}Harvard-Smithsonian Center for Astrophysics, 60 Garden
Street, MS-78, Cambridge, MA, 02138} 
\affil{\altaffilmark{3}Centro de Radioastronom\'\i a y Astrof\'\i sica, UNAM, 
Morelia, Michoac\'an, M\'exico 58089}
\affil{\altaffilmark{4}Department of Astronomy, University of Michigan, Ann 
Arbor,MI 48109}
\affil{\altaffilmark{5}NSF Astronomy \& Astrophysics Postdoctoral Fellow} 
\email{erick@astro.ugto.mx}

\begin{abstract}

CS Cha is a binary system surrounded by a circumbinary disk. We construct 
a model for the inner disk regions and compare the resulting synthetic SED
with IRS spectra of  CS Cha taken at two different epochs.  For our model we 
adopt a non-axisymmetric mass distribution from 
results of published numerical simulations of the interaction 
between a circumbinary disk and a binary system, where each star is surrounded
by a disk. In particular, we approximate the streams of mass 
from which the inner circumstellar disks accrete from the circumbinary disk. 
This structure is due to the gravitational interaction of the stars with the 
disk, in which an array of disks and streams are formed in an inner hole.
We calculate the temperature distribution
of the optically thin dust in these inner regions 
considering the variable impinging radiation from both stars and 
use the observations to estimate the mass variations in the streams. 
We find that the SEDs for both epochs can be explained with emission from an 
optically thick inner edge of the circumbinary disk and from the optically thin
streams that connect the
circumbinary disk with the two smaller circumstellar disks. To the best of our 
knowledge, this is the first time that the emission from
the optically thin material in the hole,
suggested by the theory, is tested against observations of a binary system.
\end{abstract}


\section{Introduction} 

Studying the behavior of variability in T Tauri stars surrounded by disks can lead to
a better understanding of their host stars as well as their disk structure and evolution.
With observations obtained by the Spitzer Space Telescope \citep{wer04} new detailed studies
of mid-IR variability have become possible.
Several mechanisms have been proposed to explain the  
different behavior observed in several young circumstellar 
disks. For instance, \citet{muz09} found that the mid-IR SED of LRLL~31 in
IC 348 varied on the scale of days, with a characteristic ``seesaw'' shape, 
i.e., as the emission decreased at wavelengths shortwards of $\sim 8.5 \mu$m, 
it increased at longer wavelengths. \citet{muz09} proposed that this type 
of variability reflected changes in the height of the inner disk edge.

Some infrared variability has been explained by changing the location of the 
inner disk edge.
IR variability was found by \citet{sit08} in HD 31648 and HD 
163296 in the NIR range, (specifically between $1$ and $5\mu$m ), 
where the changes could be attributed to variations in the location of the dust sublimation zone. 
Modeling under this assumption shows that the sublimation wall (boundary between the dusty and the dust-free regions of the disk) varies between $0.29AU$ and 
$0.35AU$ for HD 163296. However, any variation of the 
material in the inner region, not limited to changing the location of the inner
disk edge, could be responsible for the observed NIR variability.

Some of the intrinsic variability in circumbinary disks comes from the fact 
that the 
position of the stars is constantly changing along their orbits. Thus, the stars' irradiation of
the disk also changes, resulting in variation in the observed disk emission. 
This behavior can be 
seen in \citet{nag10}, where the predicted variation of the dust emission is 
presented for the binary system CoKu Tau$/$4.  \citet{jen07} also show that the 
emission of the binary UZ Tau E is clearly periodical. They explain this behavior with
emission from material that flows from the circumbinary disk towards the stars. 
The orbital motion of a binary system and its interaction with the circumbinary
disk has been proposed to be responsible for variability on the scale of years for UY Aur \citep{ber10}.

\citet{esp11} have presented the largest 
modeling study of mid-IR variability in disks around TTS. Most of their 
sample, composed by transitional and pre-transitional disks, 
show seesaw variability, that was explained  
as a result of variations in the height of the inner disk edge. 
However, CS Cha, the only known circumbinary disk in their sample,  
shows variations only in the 10 $\mu$m silicate emission band. 
Based on SED modeling, \citet{esp11} inferred a hole size in the disk of about $\sim 38$ AU, and this large size implies that 
the inner disk edge is too far from the stars to have a temperature 
high enough to contribute at the $\sim$ 10 $\mu$m emission. In their model, 
this 10 $\mu$m emission was produced by uniformly distributed
optically thin dust inside a radius of $\sim$ 1 AU, centered around a 
single star, and the observed variability was 
interpreted as due to variations in the mass of this optically thin dust.

CS Cha is an 
interesting laboratory to compare the observations to simulations of the interaction between 
the circumbinary disk and the binary system. 
In this paper we construct a model that takes into account results from
published numerical simulations of the circumbinary disk dynamics
\citep{gun02,gun04,art94,art96}, and aim to constrain these simulations by comparing with the variability observed in the IRS spectra presented by \citet{esp11}.
In particular, we take into account that the predicted mass distribution within the inner hole 
is non-axisymmetric; it is characterized by streams connecting the edge of the
circumbinary disk with the smaller circumstellar disks that
surround each star. We find that these streams and the circumstellar disks are 
the main contributors to the emission at 10 $\mu$m. Since this material 
has a very low density, we calculate the temperature as a function of 
grain size assuming each grain is in radiative equilibrium with the 
radiation fields of both stars. 

The paper is organized as follows.
\S~\ref{sec-observations} summarizes the observations, already 
described in \citet{esp11}. 
The model is described in
section~\ref{sec-code}, with particular detail given to how we estimate the 
spatial distribution of dust and calculate its temperature.
The model input parameters, both those 
taken to be fixed and those adopted as free parameters, 
are described in \S~\ref{sec-parameters}.
\S~\ref{sec-results} describes the results and, finally, 
\S~\ref{sec-conclusions} presents the summary and conclusions of the paper.

\section{Observations}
\label{sec-observations}

CS Cha was observed at 3 different times. The first spectrum was 
obtained
through the IRS Guaranteed-Time Observations (GTO), on July 11th 2005, and
was previously presented in \citet{esp07a}, \citet{kim09}, \citet{fur09}, and 
\citet{man11}. The other two spectra were obtained in 
GO 50403 Spitzer program (PI:Espaillat) on June 1th and 8th in 2008 and were 
first presented in \citet{esp11}. The two GO observations do not show
large differences between them. Therefore, in this work we only present the IRS
spectrum taken on June 1, 2008.

\citet{luh04} found an spectral type of $K6$ for CSCha; at that time, the
binarity was not established. 
Using this information, \citet{esp07a} found a value for the 
extinction, $A_{V}=0.8$: matching the V-I color of the star to that of a main 
sequence standard from \citet{ken95}. They take
this value to deredden the Spitzer spectrum using the Mathis reddening law
\citep{mat90}.  We adopt the same SED in this work.
A distance of $d=160pc$ is adopted for this system \citep{whi97}.

\begin{figure}
\epsscale{1.0}
\rotatebox{0}{\plotone{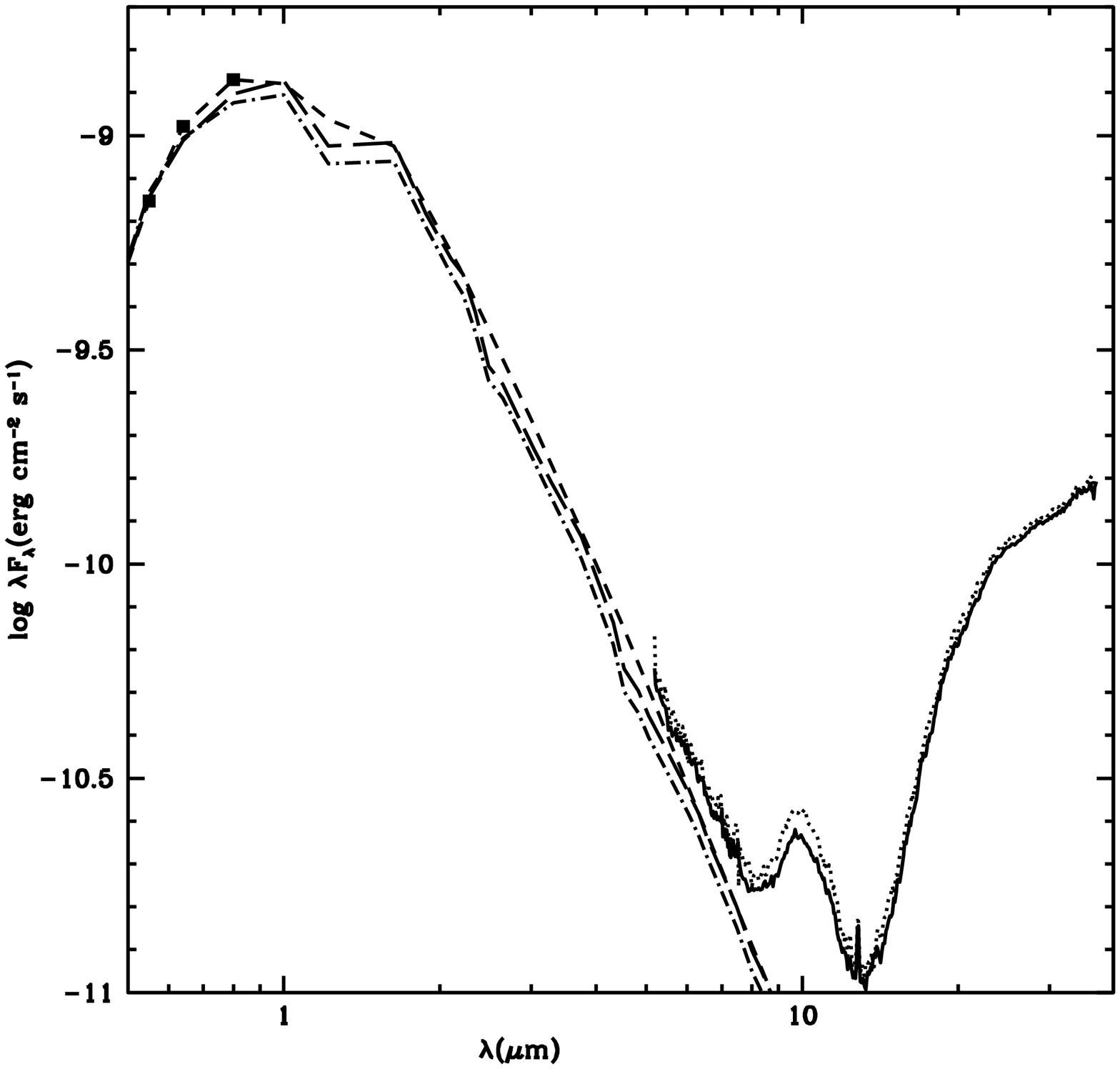}}
\caption{Comparison between the stellar spectrum used in \citet{esp11}
and in this work. For reference the observations of CS Cha are represented with
solid and pointed lines. The optical data is shown as solid squares 
\citep{gau92}. The spectrum for the isolated star used in \citet{esp11} is shown as a dashed line. The long dashed and point dashed lines
represent the combined spectrum 
obtained from both stars for Case 1 and 2, respectively. }
\label{fig-SEDstars}
\end{figure}

\citet{gue07} found CS Cha to be a binary. They did not report the
precise orbital parameters, but we use their results as reference. 
They found a fit with an orbital period of $2482$ days with a 
companion of a mass larger or equal than $0.1M_{\odot}$.
\citet{esp11} take a star of $0.91M_{\odot}$ in their single star 
model. Using the previous restriction on the secondary mass and the value of 
the primary mass in the single star model, we consider two cases
for the binary system: Case 1 is a 
set of stars with masses of $0.8M_{\odot}$ and $0.3M_{\odot}$, and Case 2
corresponds to $0.9M_{\odot}$ and $0.1M_{\odot}$ stars. For an estimated age of
$2Myr$ \citep{luh04} for the system and the masses of the stars, we use the
pre-main sequence isochrones of \citet{sie00} to find the
radius $R_{\star}$ and the surface temperature $T_{\star}$ for each star.
Our choice is justified by fitting the observed flux at low wavelengths, where 
the contribution of the disk is very small. We decided not to increase the 
secondary mass value further, because in that case, the spectral type of the
primary would be far from the one derived by \citet{luh04}. In order to argue
this, we assume that the primary spectral type is close to the estimate
which takes into account just one star \citep{luh04}.
A comparison between the SED of the single star used in
\citet{esp11} and the addition of both star's spectra for Case 1 and 
2 is shown in Figure~\ref{fig-SEDstars}. Finally, \citet{esp07a} 
estimate a mass accretion rate of 
$1.2\times 10^{-8}M_{\odot}yr^{-1}$ using the UV excess on one star, which we 
distribute between both stars (see \S~\ref{sec-fixed}). 
The parameters for both combinations of stars (Case 1 and 2) are presented 
in Table~\ref{table-stars}.

\begin{deluxetable}{ccccccccc}
\rotate
\tablewidth{0pt}
\tablecaption{Stellar parameters for the tested cases\label{table-stars}}
\tablehead{
\colhead{Case }   &
\colhead{$M_{\star1}(M_{\odot})$}      &
\colhead{$T_{\star1}(K)$}     & 
\colhead{$R_{\star1}(R_{\odot})$}      &
\colhead{$\dot{M}_{\star1}(M_{\odot}yr^{-1})$}    &
\colhead{$M_{\star2}(M_{\odot})$}      &
\colhead{$T_{\star2}(K)$}     & 
\colhead{$R_{\star2}(R_{\odot})$}      &
\colhead{$\dot{M}_{\star2}(M_{\odot}yr^{-1})$}}
\startdata
1 & 0.8 & 4059 & 1.74 & $0.4E-8$ & 0.3 & 3313 & 1.29 & $0.8E-8$  \\
2 & 0.9 & 4204 & 1.8 & $0.4E-8$ & 0.1 & 2998 & 0.85 & $0.8E-8$ 
\enddata
\end{deluxetable}

\section{The model}
\label{sec-code}

The basic model for CS Cha is similar to the one described in \citet{nag10}. 
In summary, it includes the emission of the inner wall of a 
circumbinary disk illuminated by the two stars of the binary system.  It also 
includes the emission of optically thin dust in the inner ``hole" of the disk. 
As noted in \citet{esp07a,esp11}, the size of the hole is too large to
be explained solely by the interaction of the binary with the disk. 
The outer regions of the circumbinary disk does 
not contribute at the wavelengths observed (mid-infrared), and for this reason,
it is not included here \citep{dal05,esp07a}. In principle, the 
emission of the wall is intrinsically variable with a period given by the 
orbital period of the binary system.  This is because the two stars are 
changing their 
distance relative to different points on the wall.  In addition, the wall has a
different 
projected surface in the plane of the sky.  However, in the case of CS Cha, 
the wall is located too far from the stars, according to the SED modeling 
conducted by \citet{esp11}. Using models incorporating two stars 
illuminating the wall, we found that the predicted difference in the wall 
emission is too small 
to explain the observations ($1\%$ at the $10\mu$m peak). 
As \citet{esp11}, we propose that the mid-IR variability of CS Cha is
due to variations in mass and illumination of 
optically thin dust in the circumbinary disk inner hole.  We adopt a spatial 
distribution for this dust which it is based on results of numerical simulations
in the literature.

\subsection{Outer wall}
\label{sec-wallcode}

We assume that the edge of the circumbinary disk is a 
vertical cylindrical wall illuminated by the 
stars. To the best of our knowledge, there are no 
calculations on the shape of the wall in circumbinary disks. 
For accretion disks around single stars, it has been 
proposed that the inner wall is curved, because the sublimation 
temperature of the dust grains depends on density \citep{ise05} 
and on grain size \citep{tan07}. However, in the case
of a circumbinary disk, the inner disk truncation is produced by 
dynamical effects and not dust sublimation.  
For the case of disks with embedded planets, there are detailed calculations by
\citet{var04} and \citet{cri06} of the surface
density profile of the gap, that in principle could be used to define a
shape of the inner wall. However, the cases discussed in these papers 
correspond to low mass companions (planets) and it is not clear 
how one can scale this to binary systems.  In addition,   
the density profile depends on viscosity, the disk aspect ratio, etc., 
which are unknown properties for CS Cha. 

We assume that the dust in each surface element of the inner edge of the 
circumbinary disk 
is in radiative equilibrium with the impinging radiation 
field from the stars. 
The flux received at each surface element of the wall is the addition
of the contributions of both stars, taking into account the geometrical dilution
of the stellar radiative flux, for the different distances of the stars 
at different orbital positions. Also, we assume that each star is 
accreting from a small circumstellar accretion disk, that receives 
mass from the circumbinary disk \citep{gun02}. 
It is usually assumed that for T Tauri stars surrounded by 
accretion disks, accretion shocks exist at the stellar surface that emit 
UV and optical radiation. We include the emission of these shocks 
as a source of irradiation \citep{muz03}. 
The luminosity liberated in these shocks is the
accretion luminosity,  $\sim L_{acc}$, and we assume an 
effective temperature for the shocks of $\sim 8000$ K \citep{cal98}.
Thus, the spectral energy distribution of each star is a combination of their 
photospheric and shock emission.

The zeroth and first moments of the radiative transfer equation are solved  
as in \citet{cal91} \citep{dal05}.
The irradiation flux interacts with the 
dust in the wall, a fraction is absorbed and another fraction 
is scattered. The scattered fraction produces a diffuse radiation field, 
which is finally absorbed at deeper layers.
The transfer equation is integrated along the radial direction (parallel to 
the disk midplane) for each surface element of the disk inner wall and a 
temperature is calculated assuming radiative equilibrium.  The temperature 
is a function of radial optical depth.  In this calculation, we assume that the density 
of the wall at the edge of the circumbinary disk is high enough to allow  
thermal equilibrium between grains of different sizes and gas. Thus, a unique
temperature describes the wall's thermal emission, for each wall location. 

For the calculation of the temperature on the surface of the wall [$T(\tau=0)$],
we use the optical properties of the dust at this temperature. The implicit
equation for this temperature is then solved with an iterative procedure.
The opacity evaluated at the wall surface temperature is used to calculate
the temperature as a function of $\tau$. 
Finally, with the temperature radial distribution
of each wall pixel, we calculate the emergent intensity, and taking into 
account the geometry of the wall and occultation effects, we calculate the 
contribution of the wall to the SED.

\subsection{Optically thin dust in the inner hole}
\label{sec-holecode}

The models of gap formation in circumbinary disks show a decrease 
in mass surface density between an inner hole formed due to gravitational
interactions of the binary and the disk \citep{art94} and the 
outer disk, that suggests the gap is optically thin. Note that planets immersed
in the disk are able to increase the size of the hole \citep{zhu11,dod11}. 
Observations of circumbinary disks at millimeter wavelengths, which 
resolve the inner hole: \citet{gui99} for GG Tau, \citet{gui08} for HH30, and
\citet{and10}, confirm that it has a low density 
and low optical depth. This is the case for transitional disks: \citet{hug07}
for TW Hya, \citet{hug09} for GM Aur, and \citet{and11}.  
The inner hole might have optically thin dust that contributes 
to the near and mid IR SED, e.g., TW Hya, \citep{ake11}, and 
\citep{cal05,esp07a,esp07b,esp10}. 

We consider that the distribution of material inside the hole is not 
axisymmetric (see \S~\ref{sec-structure}), following the results from 
simulations of the interaction between the binary system and the disk. 
Also, since the dust in the hole is optically thin, with very low density, we 
calculate a different temperature for each grain size, assuming radiative equilibrium
between the grain absorption of stellar and shocks radiation and emission of its own 
thermal radiation. Finally, we calculate the contribution to the SED of this 
optically thin material.

\subsubsection{Dust spatial distribution}
\label{sec-structure}

The gravitational interaction between the disk and the binary system sets 
constraints on the possible configurations of the disk material. A binary 
system is able to create a hole in the circumbinary disk \citep{lin79,art94}, 
because there is an inner region without 
stationary orbital configurations. The stable orbits for the 
particles are located very close to each star (circumstellar disks) and further
out, in trajectories around both stars (circumbinary disk; 
\citet{pic05,nag08}). As one would expect, the closer the material is 
to one of the 
stars, the less important its interaction with the other one. However, this 
configuration is perturbed 
when hydrodynamical interactions are taken into account. In this case, the 
material can lose angular momentum, and the resulting orbits end up 
connecting their 
point of origin in the circumbinary disk with each one of the circumstellar 
disks \citep{art96,gun02}. 
In order to define a structure consistent with the results of these 
hydrodynamical simulations, we model the 
system as a circumbinary disk, two circumstellar disks and 
two streams
of material, crossing the circumbinary disk hole, and connecting the outer
circumbinary disk with the inner circumstellar disks. Note that the larger hole
present in CS Cha \citep{esp07a,esp11} cannot be explained only with 
the binary, thus, one requires 
another mechanism to produce it, perhaps another stellar mass companion or a 
multiple planetary system \citep{zhu11,dod11}. 
This will change the details of the 
structure in the hole, which we will not consider here.

\citet{art96} present SPH simulations \citep{mon92} which follow
the material that
is traveling from the circumbinary disk to the stars. They note that the 
accretion is modulated in time with a periodicity given by the orbital period 
of the binary. The simple picture is that the 
material in the edge of the disk is perturbed near the apocenter of the binary,
but requires a time delay to fall to one of the stars, arriving around 
pericenter. Of course the precise evolution depends on the mass ratio and
eccentricity of the binary system \citep{art96}.
A result of this is that the time of evolution of the material through the
gap is of the order of the orbital period (P=$2482$ days). This timescale is 
smaller than a typical dust coagulation timescale ($\sim 10^{5}$ years for
$R\sim 10AU$, \citet{wei77}, \citet{bir10}), so
we assume that the grain composition and size distribution do not change
during travel.

The circumbinary material in the outskirts of the disk initially evolves 
slowly, characterized by a viscous timescale evaluated at the radius
of the circumbinary disk. Then, the material arrives to a radius defined by the
gravitational interaction of the binary system with the disk 
\citep{art94,pic05}. Starting from this radius, the matter lose
its axisymmetry and flows in two streams. The two streams are launched from
two points called the saddle points. As noted in the last paragraph, the 
characteristic time for the evolution of the material in the gap is of the 
order of the binary orbital period. Thus, when the material arrives to the 
saddle points,
it moves towards the inner circumstellar disks within a dynamical timescale.

The saddle points are the analogous to the Lagrangian points in the 3-body 
circular restricted problem in classical mechanics \citep{mur99}.
For a circular binary, one can define a rotating coordinate system where 
the stars are at rest. Thus, there is a constant of motion, the Jacobi 
constant, and specifically two 
linear Lagrangian points \citep{mur99}, that define the locus where 
material with enough value
for this constant is able to move from orbits around the binary system to
orbits connecting the circumbinary disk with circumstellar disks. 
In order to extend the analogy to an eccentric system, we note that the 
gravitational 
potential can be expanded as an infinite sum of terms. In many cases, one can 
describe the potential with only a few of them. For an
eccentric binary system, the potential can be simplified using the terms 
$(m,l)=(0,0)$ and $(m,l)=(2,1)$ of its expansion \citep{art94}. 
Moreover, this potential
rotates at a rate $\Omega_{b}/2$, where $\Omega_{b}$ is the mean angular velocity
of the binary. For this simplified case, two points can be defined 
(the corresponding Lagrangian points), which corotate with the potential.
These are the launching points for the streams mentioned in the last paragraph,
i.e. the saddle points. Analogous to the circular case, the material
that accretes from the external regions of the disk is able to cross the
radius where the saddle points are located, primarily at these locations. 
The saddle points are located at a radius $R_{s}=1.587a$, where $a$ is the 
semimajor axis. Thus, we can conclude that at larger radii, the orbital 
backbone of the material are 
orbits around the binary system \citep{pic05}. At smaller 
radii the material is located in two streams starting at the saddle points
\citep{art96,gun02} and ending in the inner disks.

\begin{figure}
\epsscale{.80}
\rotatebox{-90}{\plotone{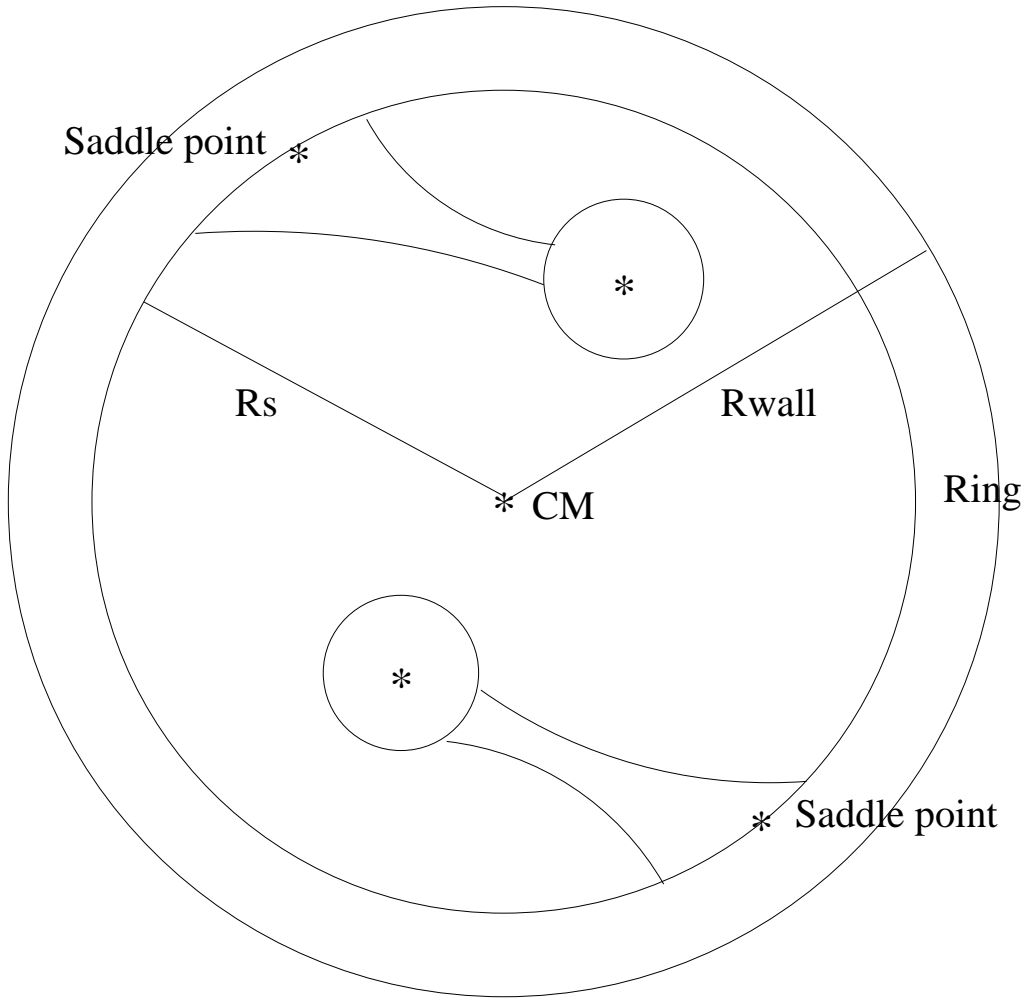}}
\caption{Sketch of the material configuration in the hole. It is not 
scaled. $R_{s}$ and $R_{wall}$ are the limits of the ring. The 
circumstellar disks and the streams are shown.}
\label{fig-structure}
\end{figure}

The configuration used in this work, which is consistent with the orbital 
restrictions given by the binary system presence, is a ring of material
between $R_{s}$ and $R_{wall}$, where $R_{wall}$ is the position of the inner
edge of the outer disk. At $R<R_{s}$, the material is located in the streams 
that connect the circumbinary and circumstellar disks. This is shown 
schematically in Figure~\ref{fig-structure}. The ring corresponds to a region
where the material evolves viscously, because this is consistent with the
gravitational disk-stars interaction as described previously. The difference
between this and the actual outer disk is that the ring is optically thin.
As we mentioned above, in the case of CS Cha, for which 
we have Spitzer mid-IR fluxes, we do not include the
contribution of the circumbinary disk to the SED, because it is too cold to
emit in this wavelength range \citep{esp07a,esp11}. Also, 
we assume that the circumstellar disks, streams and outer ring are all 
optically thin. 

\subsubsection{Dust temperature}
\label{sec-dust-temp}

We assume the optically thin dust in the inner hole is in thermal 
equilibrium with the stellar radiation. However, the density in the 
gap is so low, that thermal equilibrium between gas and dust grains is
not expected.  Therefore, each grain with a different size would have a 
different temperature. 

The critical gas density for thermal equilibrium between dust grains and 
gas can be estimated by assuming the collisional timescale is equal 
to the thermal timescale \citep{chi97,gla04}. The collisional timescale is 
given by   

\begin{equation}
 \tau_{col}={1\over n_{H}\bar{v}\pi a^{2}}, 
\end{equation}
where $n_{H}$ is the number density of the molecular hydrogen, $\bar{v}$ is the
mean velocity of a molecule and $a$ is the radius of the grain. We 
consider that the velocity
is thermal, however, it is possible that there is a turbulent component, which
amounts to a fraction of this velocity. Using the fact that the typical
velocity associated to turbulence in the $\alpha$-parametrization of the 
viscosity \citep{sha73} in protoplanetary disks \citep{haw95} is around $0.01$ 
of the sound speed, we can conclude that
including the turbulent velocity does not change our following conclusions.
The thermal timescale can be estimated as the time that a particle of dust 
takes to radiate all its thermal energy at temperature $T_{d}(a)$, 

\begin{equation}
 \tau_{ther}={kT_{d}(a)\over 4\pi a^{2}\sigma_{SB}T_{d}(a)^{4}},
\end{equation}
where $k$ is the Boltzmann constant and $\sigma_{SB}$ is the Stefan-Boltzmann
constant.

Thus, the critical density is \citep{gla04}

\begin{equation}
 n_{cr}={4\sigma_{SB}T_{d}(a)^{3}\over k\bar{v}},
\end{equation}

and substituting typical numerical values, it can be written as

\begin{equation}
 n_{cr}=(1.133\times 10^{8}cm^{-3}){T_{d}^{3}\over T_{g}^{1/2}}.
\label{eq:ncrit}
\end{equation}

where $T_g$ is the gas temperature.

Next, we would estimate the gas density associated with the optically thin 
material to compare it with this critical density.
If we assume $T_{g} \approx T_{d}$, and take $T_{d} \approx 100K$ as 
characteristic of the streams, then $n_{cr}=3.34\times 10^{-11}gcm^{-3}$. 
Note that a change in $T_{g}$ of an order of magnitude corresponds to a factor
of $3$ change in $n_{cr}$ (see eq.~[\ref{eq:ncrit}]), which does not modify
our conclusions.

In order to get an order of magnitude for the density in the streams, we note 
that the area covered by the streams is around $0.1\pi R_{s}^2$ \citep{art96}. 
We assume that
the material falls from the inner edge of the ring towards the
star in a few times the free fall time, which is $1.5$ years. This is 
consistent with the
fact that in \citet{art96}, the streams change on the order of
an orbital period of the binary, $T_{orb}\sim 7$ years. Thus, according to a
mass accretion rate $\dot{M}=10^{-8}M_{\odot}yr^{-1}$ \citep{esp07a}, 
the mass in the streams should be around $7\times 10^{-8}M_{\odot}$.
Note that in the 
hydrodynamical simulations of a disk around a binary system, \citet{gun02}
assume a typical height in terms of the radius $R$ in the disk, 
without dependence on the azimuthal angle, which is given by 

\begin{equation}
 H(R)=(\Sigma_{i=1,2}{GM_{i}\over c_{s}^{2}|R-R_{i}|^{3}})^{-1/2},
\end{equation}

where $G$ is the gravitational constant, $M_{i}$ is the mass of each of the 
stars, $c_{s}$ is the sound speed, and finally $R_{i}$ is the distance of each 
star with respect to the origin. For Case 1 and using a typical temperature of
$100\,K$ for molecular hydrogen and $|R-R_{i}|=R_{s}$, $H(R)=0.049R_{s}$.  
Thus, a typical value for $H$ is $0.1R_{s}$, which we consider here.  
Differences in $H$ do not change the following conclusions.

Given the mass in the streams, the area covered and the height calculated
previously, the characteristic density in the streams is around 
$n_{str}=10^{-14}gcm^{-3}$, which means that $n_{str} << n_{cr}$. This implies 
that the gas and the dust grains are not in thermal equilibrium, 
and since the absorption coefficient of the grains depends 
on grain size, $a$, grains with different sizes have different temperatures, 
$T(a)$. 

\begin{figure}
\epsscale{1.0}
\rotatebox{0}{\plotone{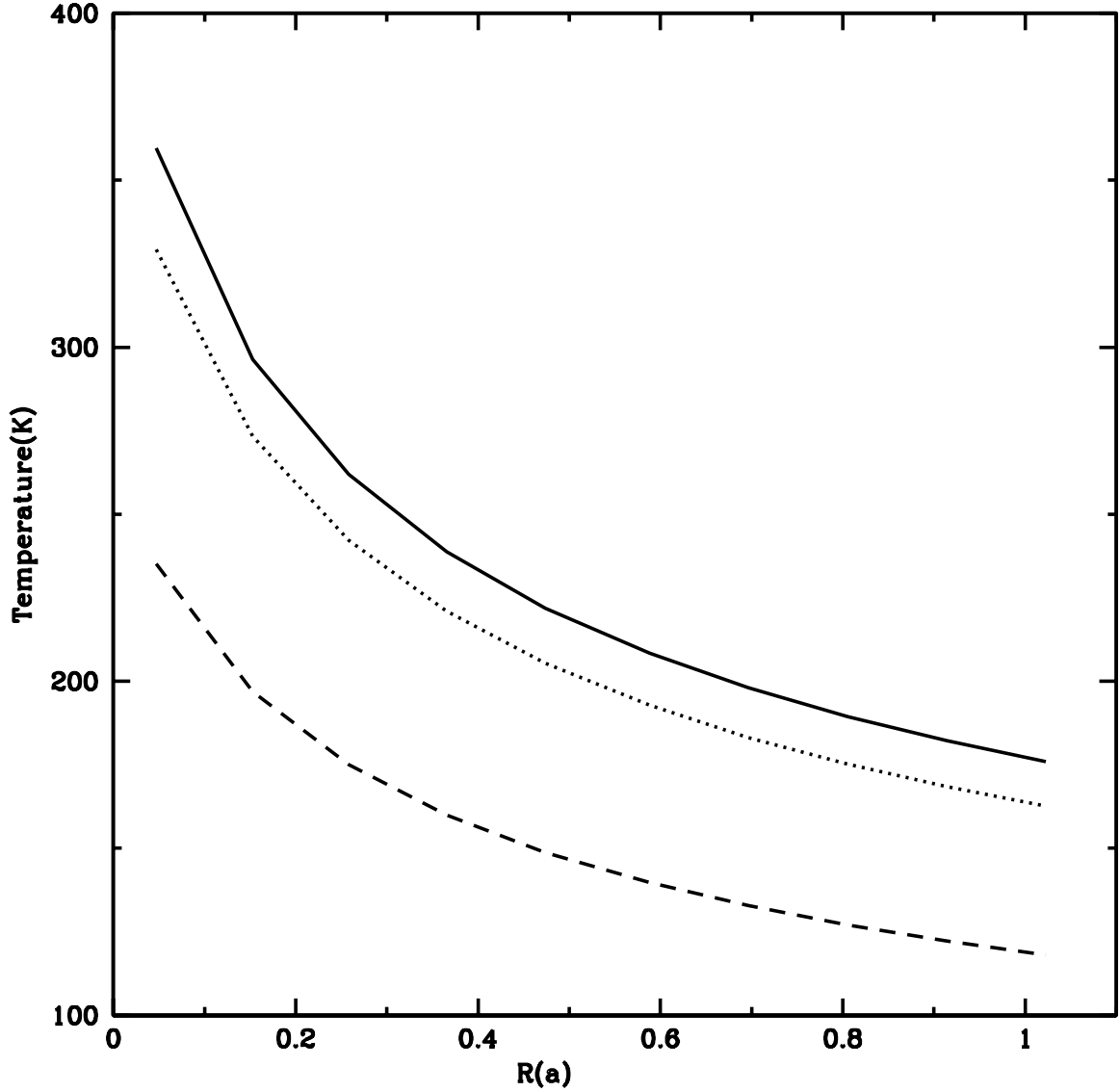}}
\caption{Temperature along the stream connecting the primary star and the 
circumbinary edge, for the fit of the GO observation for Case 1. Small grains
(solid line) correspond to a radius of $a=0.005\mu$m, large grains (dashed 
line) correspond to a radius of $a=4\mu$m and the grain size distribution 
(pointed line) is a power law of the form $a^{-3.5}$,
between the small and large grains 
} 
\label{fig-temp}
\end{figure}

We divide the grain sizes in 100 bins equally spaced in $log(a)$ between 
$a_{min}=0.005\mu$m and $a_{max}=4\mu$m, distributed with a power-law profile 
with exponent $p=-3.5$. We take this value of $a_{max}$ according to the 
detailed modeling of \citet{esp11}. For each 
size, we calculate the temperature from radiative equilibrium between 
the energy of the stellar and shock radiation absorbed by the grain and the 
energy lost by radiation. 
As expected, the temperature of the smallest grains is higher than 
the temperature of the larger grains (see discussion in 
\S~\ref{sec-fit-var} and Figure~\ref{fig-temp}). 
We also estimate a mean temperature, as a unique temperature for the 
whole grain distribution, assuming the grains have a MRN power law 
size distribution, and use this temperature as a reference. This temperature 
is also calculated assuming radiative equilibrium with 
the radiation fields of the stars and their accretion shocks, but using mean 
absorption coefficients.
As expected, at the same position in space, the temperature of the smallest 
grains is higher and the temperature of the larger grains is lower than this 
mean temperature, due to a higher heating efficiency. 
Figure~\ref{fig-temp} illustrates this by showing temperature vs distance in 
the primary's stream (Case 1, GO observations), for the smallest grain
($0.005\mu$m), the largest grain ($4\mu$m) and for a power law distribution of 
grain sizes between $0.005\mu$m and $4\mu$m.  

Remember that the estimation of the temperature is done for all the locations 
where there is material and in each case, the temperature of each grain size
is calculated. The grain composition adopted here is amorphous olivine, 
consistent with \citet{esp11}, where $95\%$ is olivine and $5\%$ is 
crystalline silicates. In addition, we include organics and troilite. The 
abundances are $\zeta_{sil}=0.0034$, $\zeta_{org}=0.001$, and 
$\zeta_{troi}=0.000768$, for silicates, organics and troilite, respectively.

\section{Parameters}
\label{sec-parameters}

Since there are many input model parameters, we search for 
observational and theoretical constraints for some of them.
In the next sections we describe the fixed and free model parameters.

\subsection{Fixed parameters}
\label{sec-fixed}

For the best model for CS Cha, in \citet{esp07a} and \citet{esp11}, the inner 
edge of the circumbinary 
disk which we refer to as the ``wall" is located at $R_{wall}=43AU$
and $R_{wall}=38AU$, respectively. The difference found in $R_{wall}$ for CS Cha
was due to using different dust opacities. The models of \citet{esp11}
use a different distribution of silicates with respect to the models for CS Cha
in \citet{esp07a}. In the first modeling, $60\%$ of the dust is 
silicates compared with $88\%$ in the other case. For both cases, crystalline
silicates correspond to a small amount. Here we use the most recent value, 
$R_{wall}=38AU$, because it corresponds to a better fit of the $10\mu$m band. 
At such a large distance, there is almost no difference between a model with one
star and a model with two stars, because $a<<R_{wall}$ ($a=3.703AU$ for Case 1,
$a=3.587AU$ for Case 2). Thus, it is safe to assume the wall parameters to be
the same inferred by \citet{esp11}, including the wall height 
($h=7AU$), and the inclination angle between the disk axis and the line of sight
($i=60^\circ$).

As noted in \S~\ref{sec-observations}, we fix the value estimated by
\citet{esp07a} for the mass accretion rate at 
$1.2\times 10^{-8}M_{\odot}yr^{-1}$. 
We distributed this value: 
$0.4\times 10^{-8}M_{\odot}yr^{-1}$ and $0.8\times 10^{-8}M_{\odot}yr^{-1}$, for
the primary and secondary, respectively. This is in agreement with 
\citet{art96}, where the simulations with $e=0.1$ show that the
mass accretion rate to the secondary is a factor of two larger than the mass
accretion rate to the primary star.
Notice that the accretion luminosities are an order of magnitude lower than the
star luminosity.  As a result, we expect that the precise value of the former 
does not make a substantial difference in the resulting SED. The stellar 
parameters for Case 1 and 2 are summarized in Table~\ref{table-stars}. 

The streams start at the saddle points of the simplified potential: 
\citet{art96} (see \S~\ref{sec-structure}) and end at each circumstellar
disk. For simplicity their shape is a line that connects both positions. 
Although the thickness of the streams is a free parameter, one condition 
should be held: the flows fill around $10\%$ of
the surface of the hole \citep{art96}. Following the simulations of 
\citet{art96}, we can safely assume that the density of streams
and outer ring are one tenth
of the values associated with the compact circumstellar disks. We assume that
both disks have a minimum
radius $R_{min}=0.1AU$ and a maximum radius $R_{max}=0.5AU$. As we mentioned
above, the mass accretion rate to the secondary is about two times the
value associated with the primary following \citet{art96}, see also
\citet{bat00}. For the modeling, we translate this as a
condition on the filling factor inside the hole associated to each stream.
Thus, we assume that the area occupied by the primary star stream is half the 
area of the stream reaching the secondary. The last requirement consistent 
with \citet{art96} is that the mass from the 
circumbinary disk that falls into any of the streams at apocenter is around 
twice the value at pericenter. Finally, in order to characterize the mass of 
the streams for every orbital location, we modulated the area of the streams
(equivalently the mass) with a sinusoidal function in the azimuthal angle. The 
advantage of using this information to define the stream structure is that 
their geometrical parameters are fixed. The density on each stream is
constant and it is fixed using the fit for the observed spectrum (see 
\S~\ref{sec-results}).

\subsection{Free parameters}
\label{sec-free}

The eccentricity is a parameter that is not restricted by the observations of
\citet{gue07}. \citet{art96} point out that the 
variability of the streams mass depends on the eccentricity. The amount of 
material associated with
the streams depends on the distance of the stars to the wall at apocenter, and
this strongly depends on the eccentricity. Note that for a circular binary,
the minimum distance of each star to the disk does not change for the 
axisymmetric wall, thus the 
structure is stationary and there is no variability in the amount of material
in the hole. Thus, in order to explain the variability, the system must be 
eccentric. Also,
the observed SED variations suggest significant changes in the
amount of material in the streams ($~17\%$ according to \citet{esp11}).
For the modeling, in order to look for a
best fit we take two values for the eccentricity, $e=0.1$ and $e=0.2$. 

The remaining parameter needed to fully characterize the binary system is 
$\phi$. It is 
the angular position of the stars along the orbit at the time we are observing 
the system: $\phi=0.5\pi$ corresponds to pericenter
and $\phi=1.5\pi$ to apocenter.
The values we try in order to find the best fit for the earlier observation
are $\phi_{1}=(0.0,0.5,1.0,1.5)\pi$. The second epoch corresponds to 
$\phi_{2}$, constrained by the time elapsed between 
the two observations, i.e., 1065 days. 
We solve the equations of motion of the binary system, looking for 
configurations separated by this time.

Finally, the dust mass surface density for the region assigned to the streams
is a free parameter. The density is increased from zero until the flux level in the
SED is reached.

\section{Results}
\label{sec-results} 

We calculate models at some representative points on the parameter space, as
given in \S~\ref{sec-free}.
For each pair of central stars (case 1 and case 2) and 
each pair of values $(e,\phi_{1,2})$, we calculate a SED 
adding the contributions of the stars, the wall and the optically thin dust, 
scaled by the dust mass surface density. The synthetic and observed SEDs are 
subtracted, and we calculate their corresponding $\chi^2$.  The mass surface 
density in each epoch is taken to be the one that minimizes $\chi^2$.

\subsection{Fitting the SED}

The fit is done independently for both epochs. In order to find the best 
fitting model, we search 
for the model that minimizes $\chi^2$. Because this is a small set of the 
parameter space, we cannot claim that we find a unique model. Due to 
the large number of assumptions and unknowns (for example, the exact 
configuration of the streams), degeneracy of the SED fitting is 
expected. Thus, one can only claim that we find models consistent with the 
observations.

\begin{figure}
\epsscale{1.0}
\rotatebox{0}{\plotone{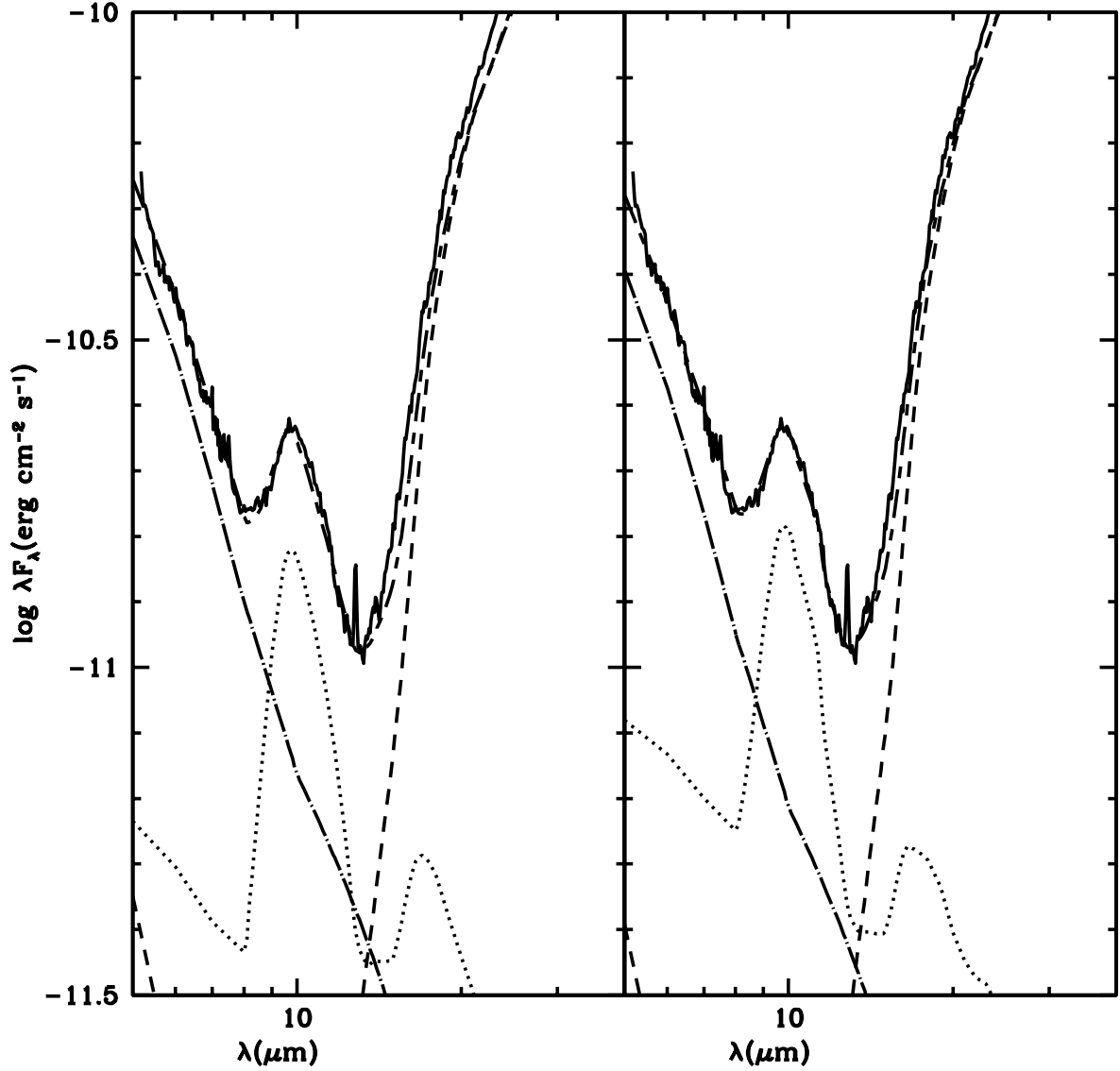}}
\caption{The best fit for CS Cha for Cases 1 (left panel) and 2 (right panel) 
(see Table~\ref{table-stars}) for the GTO observation on 2005.
The solid line is the SED observation. The pointed line is the hole SED for the
model. The dashed line corresponds to the wall flux. 
The addition of both stellar SEDs are represented with a point-long-dashed 
line.
Finally, the small-long-dashed line is the total flux for the models.}
\label{fig-CSCha-fitT}
\end{figure}

\begin{figure}
\epsscale{1.0}
\rotatebox{0}{\plotone{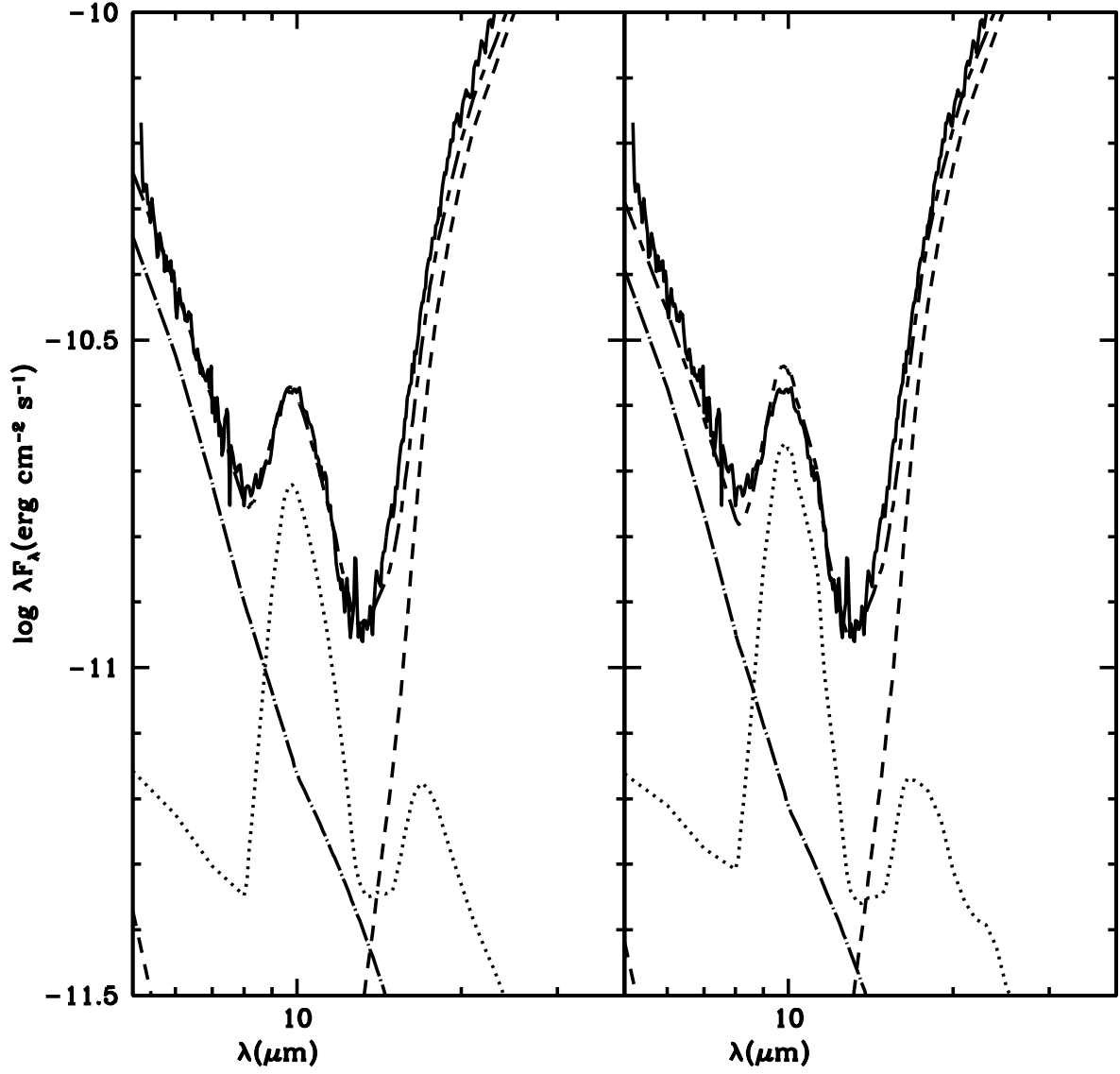}}
\caption{The best fit for CS Cha for Cases 1 (left panel) and 2 (right panel)
(see Table~\ref{table-stars}) for the GO observation on 2008.
The lines have the same meaning as Figure~\ref{fig-CSCha-fitT}.}
\label{fig-CSCha-fitTa}
\end{figure}

\begin{figure}
\epsscale{0.9}
\scalebox{0.5}[1.0]{
\rotatebox{0}{\plotone{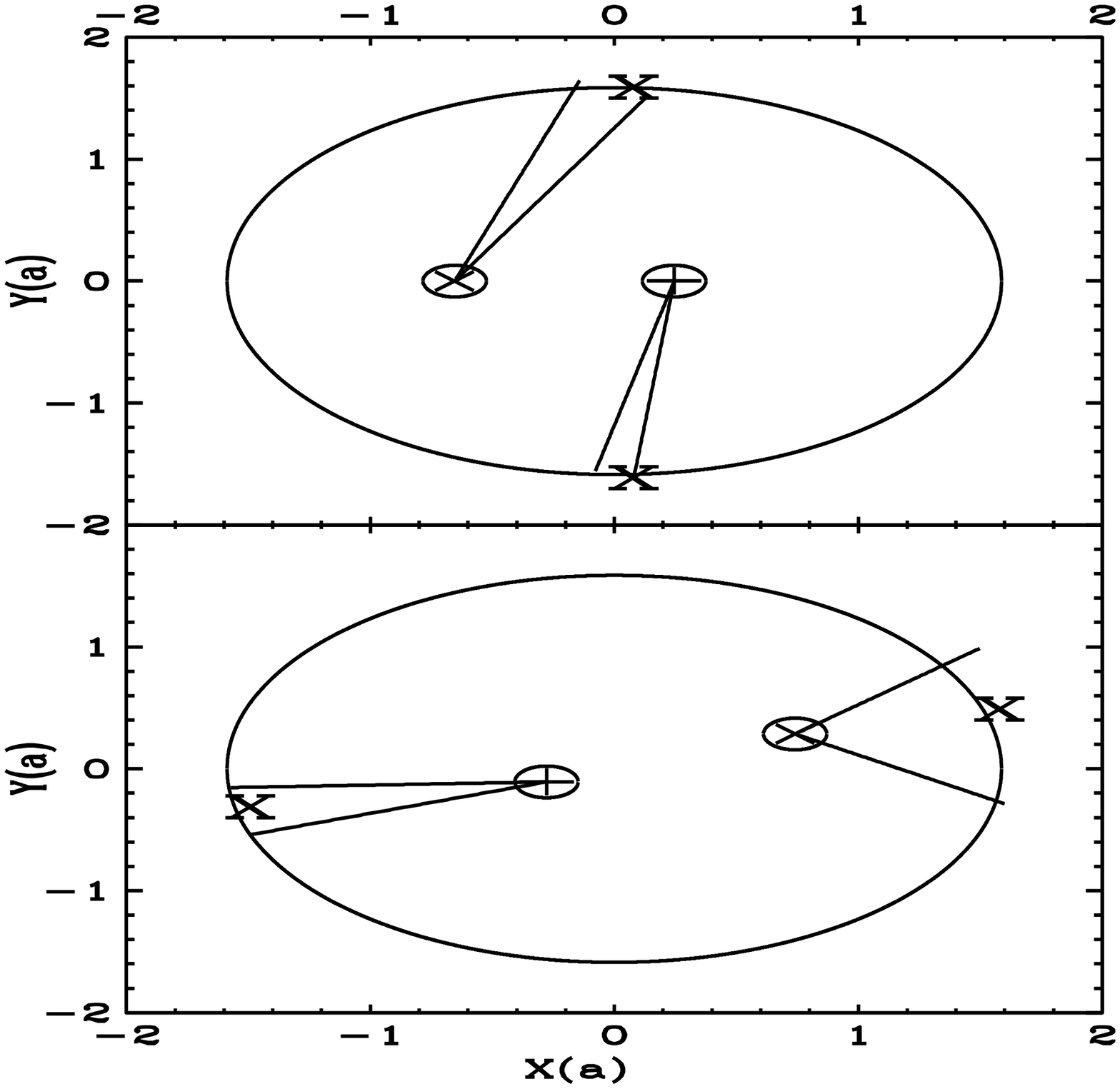}}}
\caption{Diagram of the structure of the streams viewed pole on, for the best 
fit for the Case 1 (see Table~\ref{table-stars}) for the GTO observation (upper
plot) and the GO observation (lower plot). The observer is at an inclination 
of $60^\circ$. The crosses (+) represent the primary star and the rotated 
crosses (x) represent the secondary star. 
The region inside both lines corresponds to 
the material associated to the streams. The circumstellar disks are drawn and
the label 'X' refers to the saddle points, where the streams are launched. 
The surrounding circle represents the inner boundary of the optically thin ring
of material. }
\label{fig-struct-fit}
\end{figure}

The fit for the GTO observation in 2005 is shown in Figure~\ref{fig-CSCha-fitT} 
for Cases 1 and 2. The fit for the GO 
observation in 2008 is shown in Figure~\ref{fig-CSCha-fitTa} also for both 
cases. The parameters for the best
fit model for Cases 1 and 2 are the same, this is $e=0.1$ and 
$\phi=(0.5,1.38)\pi$. 
The shape of the streams' structure for the fit is shown in 
Figure~\ref{fig-struct-fit} for Case 1, compared to the GTO and GO
observation of CS Cha. The streams for Case 2 are
almost identical to the ones of Case 1, thus, we do not show them here. 

\subsection{Fitting the observed variability}
\label{sec-fit-var}

\begin{figure}
\epsscale{1.0}
\rotatebox{0}{\plotone{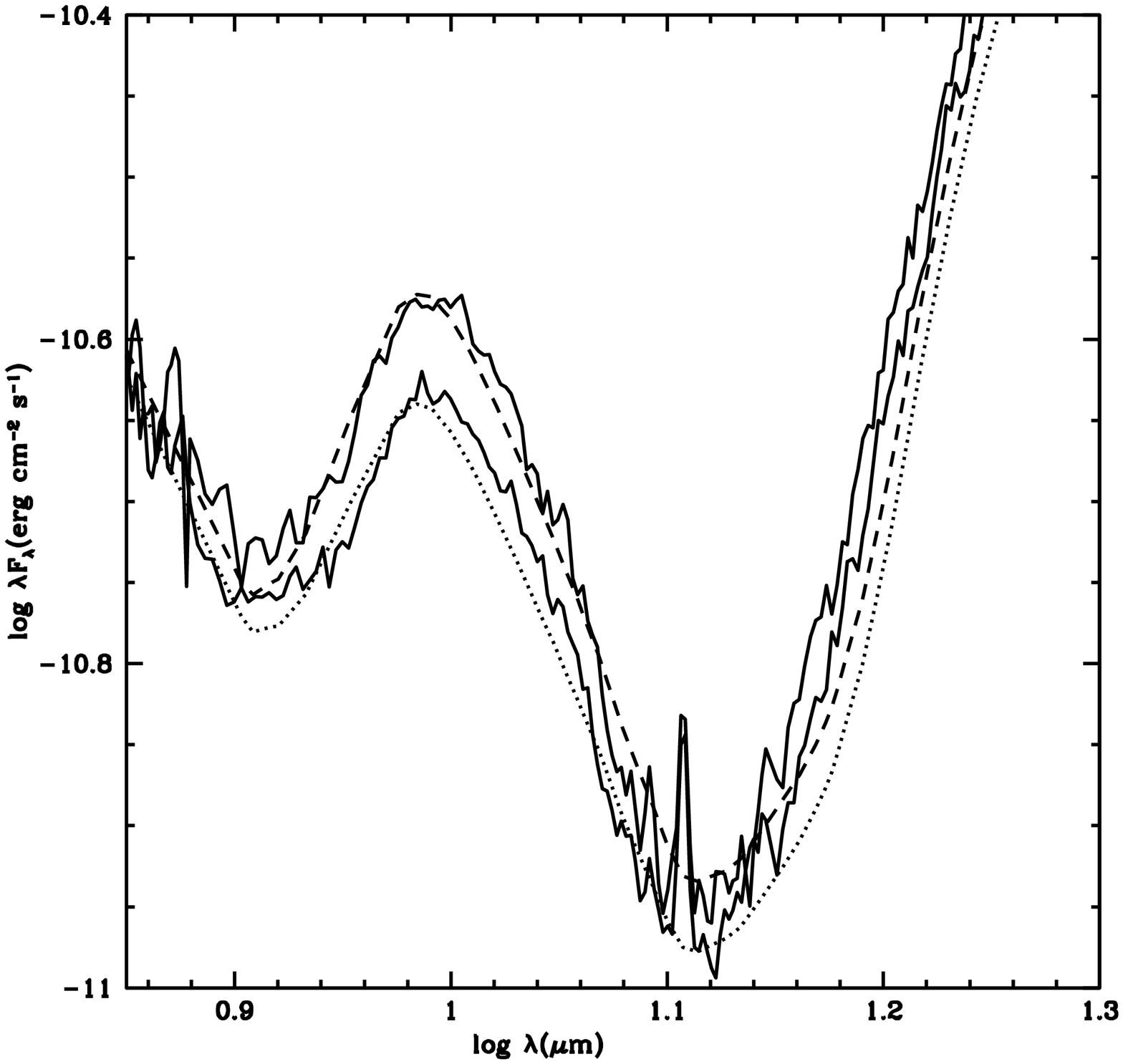}}
\caption{The modeled variability for CS Cha for the Case 1 (see 
Table~\ref{table-stars}). The solid lines represent the SED for the GTO 
observation (2005) and the GO 
observation (2008). The dashed and the pointed lines 
are the modeled SED for both epochs. These correspond to the models
shown in Figures~\ref{fig-CSCha-fitT} and \ref{fig-CSCha-fitTa}. }
\label{fig-CSCha-var1}
\end{figure}

\begin{figure}
\epsscale{1.0}
\rotatebox{0}{\plotone{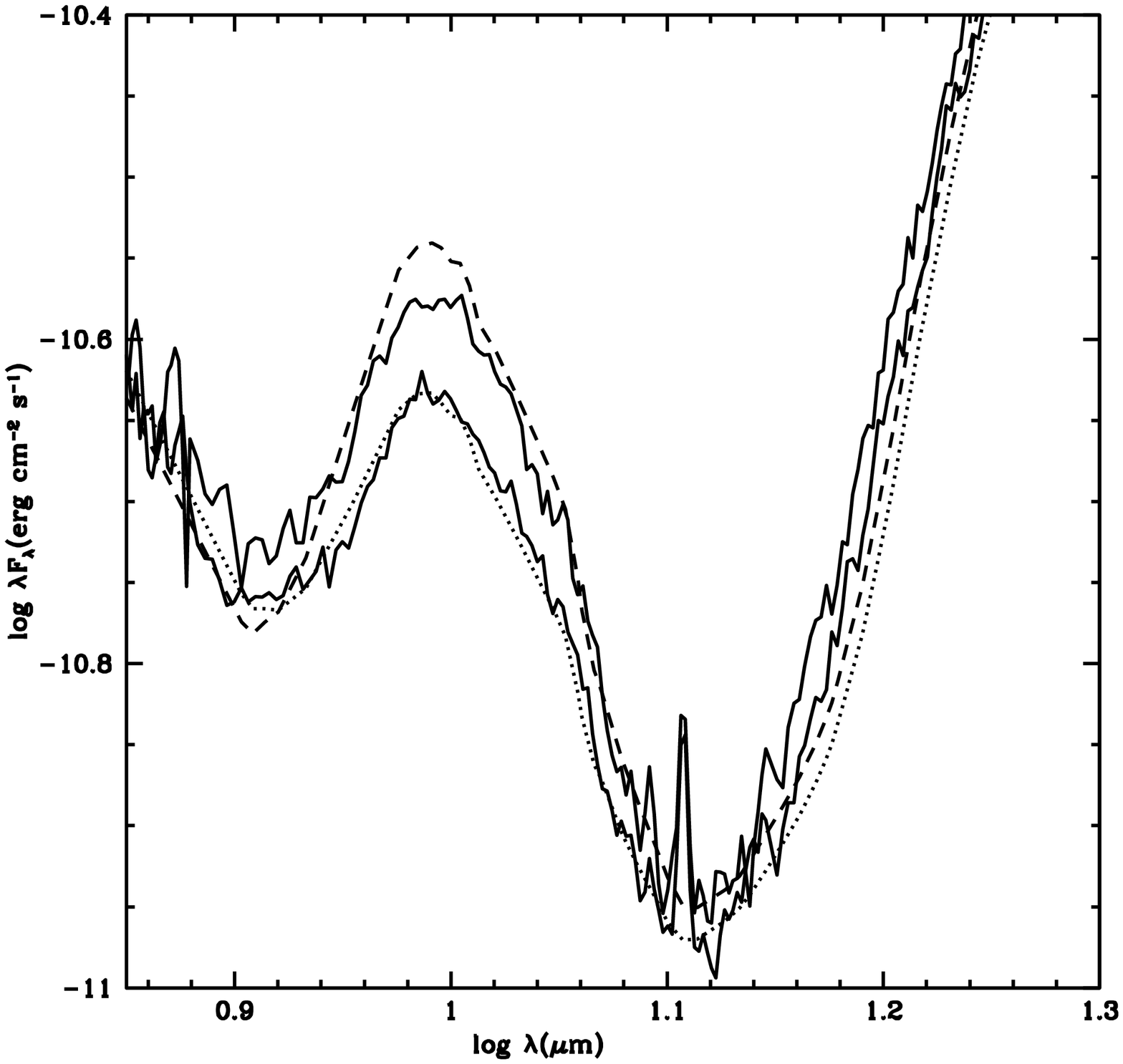}}
\caption{The modeled variability for CS Cha for the Case 2 (see 
Table~\ref{table-stars}). The solid lines represent the SED for the GTO 
observation (2005) and the GO 
observation (2008). The dashed and the pointed lines 
are the modeled SED for both epochs. These correspond to the models
shown in Figures~\ref{fig-CSCha-fitT} and \ref{fig-CSCha-fitTa}. }
\label{fig-CSCha-var2}
\end{figure}

The fit in the last section is done for both epochs, but consistently taking
into account the time difference between the observations. Thus, the 
variability can be explained with the previous fits. In 
Figure~\ref{fig-CSCha-var1}, we show the Case 1 SED fits, for both observations,
which are previously presented in Figures~\ref{fig-CSCha-fitT} and 
\ref{fig-CSCha-fitTa}. Figure~\ref{fig-CSCha-var2} is the same but for Case 2,
showing the fits also presented in Figures~\ref{fig-CSCha-fitT} and 
\ref{fig-CSCha-fitTa}. We note 
that the best fit for either models with $e=0.1$ or $e=0.2$ is $\phi_{1}=0.5$.
In order to check this tendency for every $e$, we ran models for $e=0.5$ and 
$e=0.9$. The tendency is corroborated, but the best models according to 
$\chi^2$ are worst for larger $e$. We ran models for $e=0.1$ for
other values of $\phi_{1}$, such that the whole set is 
$\phi_{1}=(0.0,0.25,0.5,0.75,1.0,1.25,1.5,1.75)$. Even for this larger set, the
best fit is found for $\phi=0.5$. 

Given the observational constraints and the properties we have fixed for 
the inner wall based on our previous modeling \citep{esp11}
we find that any eccentricity that satisfies  $e > 0$ and $\phi_{1}$ around 
$0.5$, produces a configuration
consistent with the observed SEDs in the two epochs. This means that the
eccentricity $e$ is not constrained by our models and it appears that only 
observations of the detailed orbit of the binary system would allow to quantify 
this parameter.

The SED of the present models (for both, Case 1 and Case 2) are similar to the 
SED of a single star model presented by \citet{esp11}. In the single 
star model the SED variation is related to a change in the mass of the 
optically thin dust in the inner hole. 
In the present model with two central stars, we also consider a change in 
dust mass, but besides this there is a source of 
variability given by the change in illumination produced by the stars, which 
is absent in a single star model.

\begin{figure}
\epsscale{0.6}
\rotatebox{-90}{\plotone{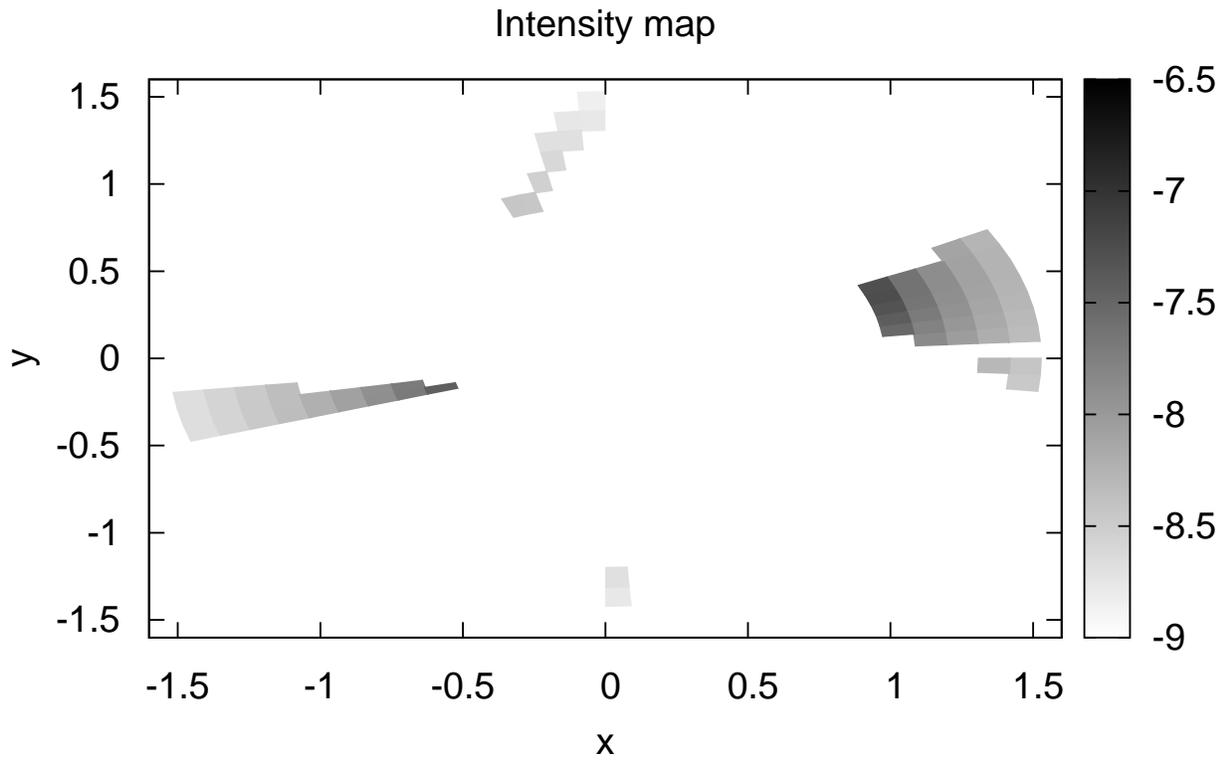}}
\caption{Intensity map at $\lambda=10\mu$m for the Case 1 fit. The gray scale 
shows the log of the intensity, which is presented at the right with units 
$erg\,cm^{-2} s^{-1} ster^{-1}$. The 
coordinates x and y are given in units of $a$. We made a 
superposition of both configuration (see Figure~\ref{fig-struct-fit}).}
\label{fig-intensity}
\end{figure}

As we show, the variability can be explained by the differences for each 
orbital configuration of the binary system, in the illumination of the material
in the circumstellar disks and streams. Also it is important the amount and
the geometrical configuration of the dust in terms of the orbital configuration.
A visualization of this effect is given in Figure~\ref{fig-intensity}, where
for the Case 1 fit, the intensity for a wavelength of $10\mu$m is presented.
One can clearly see that the contribution to the emission is larger for the
configuration associated to the GO observation (Figure~\ref{fig-struct-fit}).
Notice that in Figure~\ref{fig-intensity}, the variation in area for the 
streams means a larger amount of mass, and the gray scale corresponds to changes
in the illumination.

The associated dust mass in circumstellar disks and streams is 
around $10^{-12}M_{\odot}$, of the same order of the estimate by \citet{esp11}, 
where the material was located only in one circumstellar disk. 
The dust surface density on each stream is uniform and can be calculated 
with the total mass of dust and the area associated to each stream. In 
Table~\ref{table-density}, we present the dust surface density for the primary 
and the secondary streams, for Case 1 and 2, and for the configurations 
associated with both epochs.
In order to estimate from the model a mass accretion rate we
do the following. We assume that the gas attached to the dust in the disks and
streams is responsible for the mass accretion rate estimated in \citet{esp07a}.
Using the dust abundances, the mass of gas is of the order of
$10^{-10}M_{\odot}$, if we now use a free fall time from the saddle point as the
fastest time for the accretion, the mass accretion rate is around 
$10^{-10}M_{\odot}yr^{-1}$. This value is two orders of magnitude smaller than the
estimate in \citet{esp07a}. A possible explanation for this is that
the dust to gas ratio is lower than usually expected. 
A way to do this is through a filtering effect in the inner edge of the
circumbinary disk. \citet{ric06} note that the pressure gradient in this
location act as a filter, only allowing small dust grains to trespass it,
holding the larger ones. However, the typical critical size is around $10\mu$m,
which is larger than the maximum value taken in our work, $a_{max}=4\mu$m.
Another idea is that
the mass accretion rate inferred from U-band excess has uncertainties due to the
estimation process in itself and due to the fact that the system is a 
binary. An improvement of this estimate requires a detailed description of the
evolution of the particles (solid and gaseous) since they leave the inner edge 
of the circumbinary disk to finally accrete at one of the stars 
\citep{gun02}. Noteworthy, an U-band excess is not 
always an indicative of accretion, because one has to check the shape of the 
lines profiles to safely assume that the system is accreting \citep{ing11}. 
However, a complete study able to elucidate the dynamics of the hole
material is beyond the scope of this paper. 

\begin{deluxetable}{cccc}
\rotate
\tablewidth{0pt}
\tablecaption{Orbital parameters for the cases tested\label{table-models}}
\tablehead{
\colhead{Case }   &
\colhead{$a(AU)$}      &
\colhead{$e$}     & 
\colhead{$\phi(\pi)$}}
\startdata
1 & 3.703 & 0.1 & 0.0 \\
... & ... & ... & 0.5 \\
... & ... & ... & 1.0 \\
... & ... & ... & 1.5 \\
... & ... & 0.2 & 0.0 \\
... & ... & ... & 0.5 \\
... & ... & ... & 1.0 \\
... & ... & ... & 1.5 \\
2 & 3.587 & 0.1 & 0.0 \\
... & ... & ... & 0.5 \\
... & ... & ... & 1.0 \\
... & ... & ... & 1.5 \\
... & ... & 0.2 & 0.0 \\
... & ... & ... & 0.5 \\
... & ... & ... & 1.0 \\
... & ... & ... & 1.5 
\enddata
\end{deluxetable}

\begin{deluxetable}{cccc}
\rotate
\tablewidth{0pt}
\tablecaption{Surface density ($\Sigma$) of the streams\label{table-density}}
\tablehead{
\colhead{Case }   &
\colhead{$Epoch$}      &
\colhead{$Stream\,to$}     & 
\colhead{$\Sigma (gr\,cm^{-2})$}}
\startdata
1   & GTO  & primary   & 1.76E-5 \\
... & GTO  & secondary & 1.92E-5 \\
... & GO   & primary   & 1.58E-5 \\
... & GO   & secondary & 1.45E-5 \\
2   & GTO  & primary   & 1.91E-5 \\
... & GTO  & secondary & 4.05E-5 \\
... & GO   & primary   & 2.85E-5 \\
... & GO   & secondary & 1.43E-5 \\
\enddata
\end{deluxetable}

In previous works \citep{dal05,nag10,esp11}, the dust temperature in the hole 
and the emissivity are 
grain-size-mean values calculated using an opacity, given by the average 
opacity over the grain size distribution. In here, since the dust in the 
hole has such a low density and cannot be considered in thermal 
equilibrium with the gas (see \S~\ref{sec-dust-temp}, \citep{chi97,gla04}, we 
calculate a dust temperature and emissivity 
as a function of grain radii. For the sake of evaluating the differences 
between both approaches, we calculate a model with the grain-size-mean 
opacity to fit the SED. We find that the resulting SEDs are very 
similar, but for the model with the grain-size-mean opacity assumption, the 
mass of the optically thin material is 19\% of the mass of the models where 
no thermal equilibrium is assumed. This difference between both approaches 
means that it is worthed to do the proper estimate.

\section{Summary and conclusions}
\label{sec-conclusions}

We have shown that the Spitzer IRS spectrum of the binary CS Cha is consistent 
with the emission from the inner disk structure generated by the double system. 
This is a step forward with respect to the modeling in \citet{esp11},
because we include the binary system and more details about the hole dust 
configuration. The presence of a binary
means that the disk is gravitationally truncated, forming an inner edge (wall) 
which is directly illuminated by the stars. This would be a good candidate to 
follow up for mm imaging to confirm the hole size. 

As already discussed by \citet{esp11}, the mid-IR variability of CS 
Cha is restricted to the $10\mu$m silicate band. This implies that it is 
not produced by the variable illumination of the wall, which has a large 
radius as inferred from SED fitting,  but by variations in the emission 
of optically thin dust in the inner hole of the circumbinary disk.
We adopt a mass distribution inside the hole consistent with dynamical 
simulations (see next paragraph), i.e.,
the optically thin dust is located in a ring, two streams and two 
circumstellar disks. The mass of the streams is modulated by the orbital 
configuration of the stars, which combined with a variable illumination 
from both stars, can explain the observed variability for a reasonable 
set of parameters.

The distribution of the material in the hole is consistent with the dynamical
restrictions due to the presence of a binary system \citep{art94,pic05}. The 
material is located in a ring, streams and 
circumstellar disks (see \S~\ref{sec-structure}). This configuration is able to
explain the observed variability. 
A SED fit is done for the
optically thick regime using hydrodynamical simulations of the formation and
evolution of the hole material in a circumbinary disk in \citet{gun02} and 
\citet{gun04}. \citet{dev11} do similar simulations for the close binaries 
V4046 Sgr and DQ Tau, 
aiming to explain the shape variability of gas lines. In this case, the
density of the structure points to the optically thick regime. Thus, it is 
important to point out that for the first time (as far as we know), the SED
fitting of the mid-infrared variability, for the emission of optically thin
material inside the hole of a circumbinary disk, using a configuration 
consistent with the theory and simulations is successfully done.
 
The masses for the estimated optically thin dust are of the same order of 
magnitude than the values estimated in \citet{esp11}, however, note 
that the 
configurations of the material in the gap are quite different between both 
cases. 
Given the unknowns, we decide to calculate a small set of models in order to 
illustrate that they can explain the observed SEDs and variability, but we do
not claim this models are unique. 
However, new observations of this system will be able to increase the 
constraints, for a more extensively testing of our model. In the future, this 
model will be tested for more binary stars with
multiple observations in the infrared. Unfortunately, by now the cryogenic
mission on board the Spitzer Space Telescope is over. The ongoing warm mission 
is just able to confidently see at the smaller range of wavelengths. Thus, we 
require the launching of the James Webb Space Telescope to pursue this project 
further. The interferometer ALMA can see at a wavelength 
($330\mu$m) which is at least an order of magnitude larger than the Spitzer 
wavelength window. At this
frequency, \citet{wol05} study simulations images of a planet
embedded in a protoplanetary disk. They conclude that the planet with an 
accretion luminosity of $10^{-5}L\star$ is able to perturb the disk in order to
be detected by ALMA. In our case, the ratio of the emission from the stellar 
peak and the estimated emission at $330\mu$m is around $10^{6}$, thus, we 
think that the streams can barely be seen with ALMA.

\acknowledgments
E.N. thanks the hospitality of the Department of Astronomy of the 
University of Michigan during a research stay, where part of this work was done.
C.~E.~was supported by the National Science Foundation under Award No. 0901947.
N.C. acknowledges support from NASA Origins Grant NNX08AH94G.
PD acknowledges the grant IN11209 from PAPIIT-DGAPA-UNAM and CONACYT.

%\begin{thebibliography}{1}
%\bibitem[Werner et al. (2004)]{wer04} Werner, M.W. et al. 2004, ApJS, 154, 1
%\end{thebibliography}

\bibliography{mybib}
\end{document}